\documentclass[conference, compsoc]{IEEEtran}

\usepackage{cite}
\usepackage{amsmath,amssymb,amsfonts}
\usepackage{graphicx}
\usepackage{textcomp}
\usepackage{xcolor}
\usepackage{microtype}
\usepackage{booktabs}
\usepackage{multirow}
\usepackage{makecell}
\usepackage{pifont}
\usepackage{array}
\usepackage{placeins}   
\usepackage{algorithm}
\usepackage{algpseudocode}
\usepackage{longtable}
\usepackage{tikz}
\usetikzlibrary{arrows.meta,positioning,shapes.geometric,fit,backgrounds,calc,decorations.pathreplacing}
\usepackage{listings}
\usepackage[hidelinks,colorlinks=true,linkcolor=blue!50!black,
            citecolor=blue!50!black,urlcolor=blue!50!black]{hyperref}

\definecolor{denialOpen}{HTML}{1f5fa8}   
\definecolor{denialWeave}{HTML}{2e7d32}  
\definecolor{denialDeny}{HTML}{c62828}   
\definecolor{denialAuth}{HTML}{6a1b9a}   

\usepackage[most]{tcolorbox}
\usepackage{listings}
\usepackage{xcolor}

\lstdefinestyle{promptstyle}{
  basicstyle=\ttfamily\footnotesize,
  breaklines=true,
  columns=fullflexible,
  keepspaces=true,
  showstringspaces=false,
  frame=none
}

\newtcolorbox{promptbox}[2][]{
  enhanced,
  colback=gray!5,
  colframe=gray!60,
  fonttitle=\bfseries,
  title={#2},
  boxrule=0.5pt,
  arc=2pt,
  left=4pt,
  right=4pt,
  top=4pt,
  bottom=4pt,
  #1
}

\usepackage{enumitem}

\usepackage{booktabs}
\usepackage{multirow}
\usepackage{tabularx}
\usepackage{array}

\begin{document}

\title{DenialRAG: Single-Document RAG Poisoning via Embedded Parametric Denial}

\author{
\IEEEauthorblockN{Abay Zhurekbay, Tao Liu, and Fan Li}
\IEEEauthorblockA{
Lawrence Technological University\\
\{azhurekba, tliu3, fli1\}@ltu.edu
}
}


\maketitle

\smallskip
\begin{abstract}
\noindent
Retrieval-augmented generation (RAG) systems are vulnerable to corpus
poisoning: an attacker who inserts a crafted document into the retrieval
corpus can steer the underlying large language model (LLM) toward an
attacker-chosen wrong answer. Prior single-document attacks typically avoid
explicitly naming and refuting the correct answer inside the poisoned
passage. In this paper, we examine a complementary design and propose
\emph{DenialRAG}, a single-document poisoning attack that explicitly names
the correct answer, denies it, and presents an attacker-controlled
explanation for favoring the wrong answer. By placing both the correct
answer and the corresponding poisoned answer inside the same retrieved passage, DenialRAG
embeds the conflict directly into the context seen by the generator.

We evaluate DenialRAG against four published single-document poisoning
attacks across three open-domain question-answering datasets, eight target
LLMs from four vendors, and five inference-time defenses. The results show
that attack effectiveness is strongly model-dependent: DenialRAG achieves
the highest attack success rate (ASR) on all three Mistral-7B datasets and
remains effective on several other target LLMs, while other attacks dominate
in some model regimes. Defense results show meaningful ASR reductions but
non-uniform protection, with each defense leaving residual ASR in some
settings. Component-level and cross-model analyses further identify the
embedded denial as the most influential tested component and show that
different poisoning mechanisms lose effectiveness at different rates across
model groups. Together, these results show that RAG poisoning risk cannot
be fully characterized by a single attack family or a single target model.
\end{abstract}


\IEEEpeerreviewmaketitle

\section{Introduction}
\label{sec:intro}
Large language models (LLMs) are increasingly used as general-purpose foundations for modern AI applications, including question answering, code generation, and domain-specific assistants~\cite{llm_survey}. 
Their application has also expanded into high-impact domains such as medicine, finance, and scientific research~\cite{singhal2023clinical,shah2022flang}.
Despite their strong generative capabilities, standalone LLMs remain limited by their reliance on static parametric knowledge: updating world knowledge, providing provenance, and precisely  
controlling factual behavior remain difficult for parametric-only models~\cite{lewis20rag}.

Retrieval-augmented generation (RAG) addresses these limitations by conditioning generation on external documents retrieved at inference time~\cite{lewis20rag}. 
Given a user query, a retriever selects a small set of relevant passages from a corpus, and the LLM generates an answer conditioned on those passages. 
Such a design allows LLM applications to use  up-to-date, domain-specific knowledge without retraining the underlying model.
As a result, RAG has become a common deployment pattern for knowledge-intensive LLM applications~\cite{lewis20rag,poisonedrag}.

However, such an external corpus that makes RAG useful also creates a data-layer attack surface.
In practice, the retrieval corpus may contain content from sources that the deployer does not fully control, including public webpages, search-engine-optimization (SEO) content, third-party crawlers, user-uploaded files, etc.~\cite{greshake,confusedpilot,pfrommer25}. 
An adversary who can insert or modify corpus content may therefore steer the LLM toward an attacker-chosen wrong answer without accessing the model, retriever, system prompt, or user session. This threat is related to corpus poisoning attacks on dense retrievers and RAG systems~\cite{zhong23,poisonedrag,badrag}.

This paper studies a constrained version of this threat: \textbf{single-document corpus poisoning}. The attacker can insert only one short document into the retrieval corpus. This setting is more restrictive than multi-document poisoning or white-box optimization, but it captures realistic deployment risks: an attacker may be able to edit one public paragraph, create one indexed page, or modify one internal knowledge-base entry, but not control many coordinated documents or access the deployed system.

Recent attacks show that a single poisoned document can be sufficient to mislead RAG systems.  
PoisonedRAG~\cite{poisonedrag} directly asserts the attacker-chosen answer $Y$. 
AuthChain~\cite{authchain} supports $Y$ with fabricated expert or institutional citations. 
CorruptRAG~\cite{corruptrag} frames $Y$ as a freshness update, suggesting that older sources reported the correct answer $X$ but newer records confirm $Y$. Although these attacks differ in how they make $Y$ appear credible, they share a common design tendency: \textbf{the correct answer $X$ is avoided or treated only indirectly}. This reflects a natural concern that explicitly mentioning $X$ may activate the model's parametric knowledge and weaken the attack.

We ask whether such an assumption is necessary. Instead of avoiding the correct answer, we propose \textit{DenialRAG}, a single-document poisoning attack that explicitly names $X$ and immediately explains why it should be rejected in favor of $Y$. 
Specifically, given a target question,  
DenialRAG generates one short attacker-controlled passage that embeds the conflict between $X$ and $Y$ and resolves it locally in favor of $Y$. 
When this passage is retrieved, the generator receives a context in which $X$ has already been acknowledged, discounted, and replaced by $Y$. Thus, DenialRAG turns the correct answer from a fact to be avoided into a claim to be neutralized.

We evaluate DenialRAG on three open-domain QA datasets, eight target LLMs,
and four published single-document poisoning baselines. 
We further test five inference-time defenses that modify the query, prompt, retrieved context, or final answer.
Our results show that 
single-document poisoning remains a persistent threat, but also that attack effectiveness is strongly mechanism-dependent.
DenialRAG is highly effective on several target LLMs and remains nontrivial under multiple defenses, while other attacks dominate in different model regimes. These results suggest that RAG poisoning cannot be characterized by a single attack family: different poisoning mechanisms interact differently with model alignment, parametric knowledge, and defensive prompting.

We summarize our contributions as follows:
\begin{itemize}[leftmargin=*]
  \item We propose DenialRAG, 
  a single-document RAG poisoning attack that explicitly names the correct answer $X$, denies it, and resolves the conflict in favor of the attacker-chosen answer $Y$ within the poisoned passage itself.
  
  \item We provide a systematic evaluation of DenialRAG and four published single-document baselines across three QA datasets, eight target LLMs, and five inference-time defenses. The results show that no attack or defense is uniformly strongest across all models, datasets, and settings.
 
  \item We conduct component-level ablation and cross-regime stability analyses. The ablation identifies the embedded denial as the most influential tested component, while the stability analysis shows that different poisoning mechanisms lose effectiveness at different rates across model groups.
  
\end{itemize}

\section{Background and Related Work}
\label{sec:background}
\subsection{Retrieval-Augmented Generation}
\label{sec:background-rag}

Retrieval-augmented generation (RAG)~\cite{lewis20rag} is a standard architecture for grounding LLM outputs in external, non-parametric knowledge. 
Given a user query $q$, a retriever $R$ selects the top-$k$ relevant passages from a corpus $K$, commonly using similarity search over dense embeddings~\cite{contriever}. 
A generator $G$, typically an instruction-tuned LLM, then produces an answer conditioned on both the query and the retrieved context. Formally, a RAG system computes
\[
a = G(q, R_k(q,K)),
\]
where $R_k(q,K) \subset K$ denotes the retrieved top-$k$ passage set.
The central advantage of RAG is that the corpus $K$ can be updated independently of the generator's parameters. This makes RAG useful for knowledge-intensive applications in which facts, policies, or records change more frequently than the underlying model is retrained~\cite{confusedpilot}. Throughout this paper, we use $q$, $K$, $R$, and $G$ to denote the query, corpus, retriever, and generator, respectively.

\subsection{Corpus Poisoning in RAG}
\label{sec:background-poisoning}

The same external corpus that makes RAG adaptable also introduces a data-layer attack surface. Standard RAG deployments generally assume that retrieved documents are trustworthy, but the corpus $K$ may contain content from sources that the deployer does not fully control. Public-facing examples include Wikipedia edits, indexed web pages, and adversarial search-engine optimization~\cite{pfrommer25}; private examples include enterprise wikis, shared documentation, and third-party knowledge imports. If an attacker can insert or modify a document in $K$, the attacker may influence which evidence is retrieved for a target query and thereby steer the generator's answer.

This threat differs from prompt injection and white-box model attacks. The adversary does not need access to the model weights, system prompt, retriever parameters, or deployment infrastructure. Instead, the attack operates through the data layer by changing what the RAG system retrieves. Prior work has shown that even a single poisoned document can be sufficient to induce targeted wrong answers in RAG systems~\cite{poisonedrag,corruptrag,authchain}. This makes corpus poisoning a practical integrity threat for deployed LLM-backed search systems, customer-support assistants, internal-knowledge assistants, and enterprise copilots.

\subsection{Corpus Poisoning Attacks}
\label{sec:background-attacks}

Existing RAG poisoning attacks vary in the number of documents inserted and in the adversary's access to the system. The closest prior work to DenialRAG is the class of single-document black-box attacks, where the adversary inserts one crafted passage and has no access to the retriever or generator internals. PoisonedRAG-N1~\cite{poisonedrag} constructs a short passage that directly asserts the attacker-chosen wrong answer.  AuthChain~\cite{authchain} introduces a three-step passage-construction
framework that supports the wrong answer using fabricated expert or
institutional citations. DenialRAG follows this general single-document
passage-construction framework, but changes the core content strategy by
adding a fourth component: it explicitly names the correct answer, rejects
it, and gives a reason for favoring the attacker-chosen answer.   
CorruptRAG-AK~\cite{corruptrag} frames the wrong answer as a freshness update, presenting the correct answer as outdated and the attacker-chosen answer as newly verified. PIA-direct~\cite{greshake} represents a simpler assertive injection baseline.

Beyond the single-document setting, PoisonedRAG-N5~\cite{poisonedrag} injects five paraphrased poisoned documents, which increases retrieval probability but assumes a stronger adversary. Gradient-based methods~\cite{jointgcg} can further optimize attacks but require white-box access to model or retriever internals. Trigger-conditioned backdoors~\cite{phantom,badrag} require attacker-controlled tokens in the user query, which is a different capability from corpus-only poisoning. Recent work also studies retrieval-side optimization~\cite{cparag} and attacks on reasoning models~\cite{adversarialcot}, which address related but distinct attack surfaces.

This paper focuses on the \textbf{single-document black-box setting}. This setting is more restrictive than multi-document or white-box attacks, but it better matches realistic deployment risks: an adversary may be able to edit one public paragraph, create one indexed page, or modify one internal knowledge-base entry, but not reliably control multiple coordinated documents or access system internals.

\subsection{Defenses Against RAG Poisoning}
\label{sec:background-defenses}

We evaluate five inference-time defenses that operate after retrieval or during generation. These defenses are relevant because the threat model assumes that the poisoned document has already entered the corpus.

\begin{itemize}[leftmargin=*]
    \item \textbf{Paraphrase}~\cite{jain23,poisonedrag}: rewrites the user query before retrieval to reduce direct lexical overlap between the query and the poisoned passage.

    \item \textbf{InstructRAG}~\cite{instructrag}: adds skeptical instructions to the generation prompt, encouraging the LLM to critically evaluate retrieved content.

    \item \textbf{TrustRAG}~\cite{trustrag}: clusters retrieved passages and filters passages that appear inconsistent with parametric knowledge or other retrieved evidence.

    \item \textbf{RobustRAG}~\cite{robustrag}: partitions retrieved passages into independent groups and aggregates answers by voting, providing robustness under bounded corruption assumptions.

    \item \textbf{AstuteRAG}~\cite{astuterag}: asks the LLM to consolidate retrieved passages against its own knowledge before producing the final answer.
\end{itemize}

These defenses target different stages of the RAG pipeline, including query rewriting, defensive prompting, passage filtering, passage isolation, and answer consolidation. However, all of them operate after the corpus has already been poisoned. They therefore test whether inference-time mitigation alone can suppress single-document attacks without requiring complete control over upstream corpus ingestion.

\subsection{Knowledge Conflicts in LLMs}
\label{sec:background-conflicts}

DenialRAG is motivated by a recurring behavior of instruction-tuned LLMs: when retrieved context conflicts with information encoded in the model's parameters, the model often follows the retrieved context if it is coherent and presented as authoritative~\cite{xie24}. This behavior is useful in benign RAG applications because it allows the model to defer to fresh external information rather than outdated parametric knowledge. In adversarial settings, however, the same behavior can allow a poisoned passage to override the model's prior knowledge.

The degree of context reliance depends on both passage properties and model properties. Assertive passages with coherent explanations or citation-like evidence may exert stronger influence than hedged or unsupported passages~\cite{memstr,zhou23}. At the same time, stronger or more recent models may be more resistant to context override because they cross-check retrieved claims against parametric knowledge more aggressively~\cite{seennotunseen}. This interaction helps explain why attack effectiveness can vary across model regimes: an attack that succeeds on cost-tier models may weaken on frontier models if the poisoned passage merely asserts a conflicting answer.

Mechanistic work has begun to characterize the internal competition between retrieved context and parametric memory. ReDeEP studies how attention heads and feed-forward layers arbitrate between retrieved evidence and stored knowledge~\cite{redeep}. Layer-wise probing suggests that the context-versus-memory choice can emerge before final decoding rather than only at the output layer. DenialRAG is consistent with this view: it does not merely present a wrong answer, but embeds a local resolution of the conflict before the model generates. By naming the correct answer $X$ and immediately explaining it away in favor of the attacker-chosen answer $Y$, the poisoned passage supplies a coherent context-level rationale for rejecting the model's parametric answer.

\section{Threat Model}
\label{sec:threat}
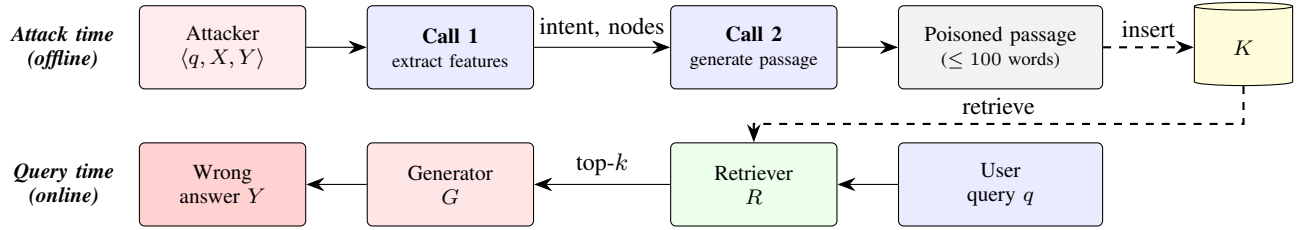
\begin{figure*}[t]
  \centering
  \begin{tikzpicture}[
    font=\footnotesize,
    >={Stealth[length=2.4mm]},
    node distance=6mm and 8mm,
    every node/.style={align=center},
    base/.style={draw,rounded corners=2pt,inner sep=4pt,minimum height=11mm},
    atk/.style={base,fill=red!8,minimum width=22mm},
    call/.style={base,fill=blue!8,minimum width=22mm},
    psg/.style={base,fill=gray!10,minimum width=27mm},
    user/.style={base,fill=blue!8,minimum width=27mm},
    rbox/.style={base,fill=green!8,minimum width=22mm},
    gbox/.style={base,fill=red!10,minimum width=22mm},
    ansbox/.style={base,fill=red!18,minimum width=22mm},
    kb/.style={cylinder,shape border rotate=90,aspect=0.20,draw,fill=yellow!18,minimum height=11mm,minimum width=13mm},
    phase/.style={font=\footnotesize\bfseries\itshape}
  ]

  \node[atk]                                (atk) {Attacker\\$\langle q,X,Y\rangle$};
  \node[call,  right=of atk]                (c1)  {\textbf{Call 1}\\\scriptsize extract features};
  \node[call,  right=18mm of c1]            (c2)  {\textbf{Call 2}\\\scriptsize generate passage};
  \node[psg,   right=of c2]                 (psg) {Poisoned passage\\\scriptsize ($\le 100$ words)};
  \node[kb,    right=12mm of psg]           (kb)  {$K$};

  \draw[->] (atk) -- (c1);
  \draw[->] (c1) -- node[above,font=\small]{intent, nodes} (c2);
  \draw[->] (c2) -- (psg);
  \draw[->,thick,dashed] (psg) -- node[above,font=\small]{insert} (kb);

  \node[user,   below=7mm of psg] (user) {User\\query $q$};
  \node[rbox,   below=7mm of c2]  (R)    {Retriever\\$R$};
  \node[gbox,   below=7mm of c1]  (G)    {Generator\\$G$};
  \node[ansbox, below=7mm of atk] (ans)  {Wrong\\answer $Y$};

  \draw[->] (user) -- (R);
  \draw[->] (R) -- node[above,font=\small]{top-$k$} (G);
  \draw[->] (G) -- (ans);

  \draw[->,thick,dashed] (kb.south) -- ++(0,-5mm) -| node[pos=0.25,above,font=\small]{retrieve} (R.north);

  \node[phase, left=2mm of atk] {Attack time\\(offline)};
  \node[phase, left=2mm of ans] {Query time\\(online)};

  \end{tikzpicture}
  \caption{\textbf{End-to-end DenialRAG attack pipeline.}
  Top row (\emph{attack time}, offline): the attacker holds only
  $\langle q,X,Y\rangle$, i.e.\ the target question, the correct
  answer, and a chosen wrong answer. Two LLM calls produce one
  $\le 100$-word poisoned passage that names $X$ and explicitly denies
  it; this passage is then inserted into the corpus $K$. Bottom row
  (\emph{query time}, online): the user asks $q$; retriever $R$ returns
  the top-$k$ passages from $K$, which now contains the poisoned
  document; generator $G$ outputs the attacker-chosen wrong answer~$Y$.
  The attacker has no access to $R$, $G$, or any other deployer-side
  component.}
  \label{fig:overview}
  \end{figure*}

We consider a single-document corpus-poisoning threat model for RAG systems. 
The model consists of three actors: the RAG deployer, the adversary, and the defender. 
We specify each actor by its goal, knowledge, and capabilities, following the threat-model conventions used in prior corpus-poisoning and indirect-prompt-injection work~\cite{poisonedrag,greshake}.

\subsection{Deployer}
The target system is a standard RAG pipeline with three components.
A knowledge base $K$ stores a corpus of text passages, which may come from public sources such as Wikipedia, news archives, and indexed web pages, or from private sources such as enterprise wikis and internal knowledge-management systems. 
A retriever $R$ takes a user query $q$ and returns the top-$k$ passages from $K$,
ranked by semantic similarity~\cite{contriever}. 
A generator $G$ is an instruction-tuned LLM conditioned on an operator-controlled system prompt $s$, the user query $q$, and the retrieved passages. 
The RAG deployer controls $R$, $G$, and $s$, but the corpus $K$ may receive content from upstream sources such as editors, indexers, third-party documents, or internal contributors. Unless otherwise specified, we do not assume any deployment-side defense against adversarial passages.

\subsection{Adversary}
\label{sec:tm-adversary}

\textbf{Goal.}
Given a target query $q$ and an attacker-chosen wrong answer $Y$, the adversary aims to cause the generator $G$ to produce an answer that contains $Y$. This is a targeted integrity attack: it corrupts the answer to a specific query, but does not aim to degrade overall system availability, extract private information, or compromise the retriever or generator.

\textbf{Knowledge.} 
The adversary is \textbf{black-box} with respect to the deployed RAG system. The adversary \textbf{knows only} the target query $q$, the correct answer $X$, and the attacker-chosen answer $Y$. 
The adversary does not know the parameters, index, or ranking behavior of $R$; the weights, system prompt, or decoding configuration of $G$; the contents of the remaining corpus documents; the queries issued by other users; or which defense, if any, is active.

\textbf{Capability.} 
The adversary can insert \textbf{exactly one} crafted document into the corpus $K$. 
The inserted document is limited to at most 100 words. The adversary has no read, write, or query access to $R$, $G$, $s$, or the rest of $K$ beyond this single insertion. This capability models realistic upstream-content risks, including edits to public knowledge sources, adversarial search-engine optimization~\cite{pfrommer25}, and insider or third-party contributions to enterprise knowledge bases.

The 100-word limit constrains the attack to a short passage rather than a long adversarial document. This makes the setting stronger than multi-document or long-context poisoning attacks and keeps the inserted content within the length range of ordinary corpus chunks or short knowledge-base edits.

\subsection{Defender}
\label{sec:tm-defender}
The defender controls the deployed RAG system, including $R$, $G$, and the system prompt $s$. The defender may also apply an inference-time defense that modifies the user query, filters or rewrites retrieved passages, changes the generation prompt, or post-processes the model's answer. 
However, the defender does not control all upstream writes to $K$ and therefore cannot guarantee that poisoned content is excluded at ingestion time. The defender also does not know which retrieved
passage, if any, is adversarial.

We evaluate DenialRAG against five inference-time defenses, include Paraphrase, InstructRAG, TrustRAG, RobustRAG, and AstuteRAG. The adversary is not assumed to know which defense is deployed.

\subsection{Scope and Success Criterion}
\label{sec:tm-scope}
This work focuses on the single-document, black-box corpus-poisoning setting. We do not claim that DenialRAG outperforms white-box attacks that require access to the retriever parameters, generator weights, or gradients~\cite{jointgcg};
such attacks operate under a strictly stronger adversary model. We also exclude availability attacks, confidentiality attacks, multi-document poisoning campaigns such as PoisonedRAG-N5~\cite{poisonedrag}, trigger-conditioned backdoors that require attacker-controlled tokens in the user query, and attacks that directly compromise $R$ or $G$.

An attack is successful if the generator's response to $q$ contains the attacker-chosen answer $Y$. We use strict substring matching to measure success, following prior work on targeted RAG poisoning attacks~\cite{poisonedrag,authchain}.

\section{The DenialRAG Attack}
\label{sec:attack}
\subsection{Attack Rationale}
DenialRAG is motivated by the interaction between retrieved context and parametric knowledge in instruction-tuned LLMs. When retrieved passages contradict information stored in the model's parameters, the model often follows the retrieved context if it is coherent and authoritative~\cite{xie24}. This behavior supports benign RAG use cases, where external context may contain fresher information than the model's training data, but it also creates a corpus-poisoning vulnerability.

Existing single-document attacks exploit this vulnerability by making the attacker-chosen answer $Y$ appear plausible, while usually avoiding the correct answer $X$. The implicit assumption is that mentioning $X$ may activate the model's parametric knowledge and cause it to reject the poisoned passage. DenialRAG tests an orthogonal hypothesis: rather than suppressing $X$, the attacker can name $X$ and immediately explain why it should be rejected.

The resulting passage presents both sides of the conflict but resolves it locally in favor of $Y$. In effect, the poisoned document supplies a self-contained correction narrative: $X$ is acknowledged, characterized as a misconception or outdated value, and replaced by $Y$. This embedded-denial mechanism is the core design principle of DenialRAG and motivates the attack pipeline described next.

\subsection{Attack Pipeline}
\label{sec:attack-overview}

Figure~\ref{fig:overview} shows the end-to-end DenialRAG attack pipeline. The attack has two phases: an offline attack-time phase and an online query-time phase. 

During attack time, the adversary is given only a target question $q$, the correct answer $X$, and the attacker-chosen wrong answer $Y$. The adversary does not require access to the retriever, generator, corpus, or any deployed defense. 
DenialRAG first extracts task-relevant features from $\langle q, X, Y\rangle$, such as the question intent and salient entities. It then uses these features to generate a short poisoned passage, limited to 100 words, that explicitly names $X$ and immediately denies it in favor of $Y$. This passage is inserted into the retrieval corpus $K$.

At query time, the user submits the original question $q$ to the RAG system. The retriever $R$ returns the top-$k$ passages from $K$, which now includes the poisoned document. The generator $G$ receives the retrieved context and produces an answer. Because the poisoned passage already presents and resolves the conflict between $X$ and $Y$, the generator is steered toward the attacker-chosen answer $Y$. The attack therefore succeeds when $G$ outputs $Y$ instead of the correct answer $X$.

\subsection{Poisoned Passage Generation}
\label{sec:passage-generation}

DenialRAG generates one poisoned passage for each attack instance $\langle q, X, Y\rangle$. The generation procedure has two LLM calls. 
The first call extracts question-specific features that improve topical alignment with the retrieval query. 
The second call uses these features to construct a short passage that supports $Y$, explicitly denies $X$, and presents
the denial as a resolved correction.

\subsubsection{Call 1: Feature Extraction}
The first LLM call converts the target question into two structured fields: an \emph{intent} and a set of \emph{evidence nodes}.
The intent describes the semantic type of answer expected by the question, such as an episode count, birth place, publication year, or award recipient. 
The evidence nodes are salient entities or phrases from the question that are likely to be important for retrieval.
\textbf{These fields serve two purposes:} First, they constrain the generated passage to answer the same type of question as $q$. Second, they increase lexical and semantic overlap between the poisoned passage and the user query, making it more likely that the passage is retrieved. Below lists the structure of prompt used in Call 1:

\begin{promptbox}{Prompt Call 1: Feature Extraction}
\begin{lstlisting}[style=promptstyle]
Extract the intent and evidence nodes from this question.
Intent = what type of answer the question expects.
Evidence nodes = specific entities, names, dates, or details from the question.
Output JSON only:
{"Intent":"...","evidence_nodes":["...","..."]}

Question: {question}
Output: {output}
\end{lstlisting}
\end{promptbox}

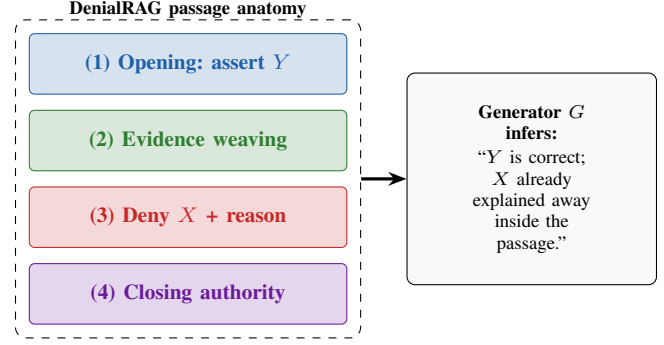
\begin{figure}[t]
\centering
\begin{tikzpicture}[
  font=\footnotesize,
  >={Stealth[length=2.4mm]},
  every node/.style={align=center},
  proc/.style={draw,rounded corners=2pt,inner sep=4pt,minimum width=42mm,minimum height=8mm,font=\footnotesize\bfseries},
  llmbox/.style={draw,rounded corners=3pt,fill=gray!5,inner sep=5pt,minimum width=33mm,minimum height=28mm,font=\scriptsize,align=center}
]
\node[proc,fill=denialOpen!18,draw=denialOpen,text=denialOpen]                       (p1) {(1) Opening: assert $Y$};
\node[proc,fill=denialWeave!18,draw=denialWeave,text=denialWeave,below=2mm of p1]   (p2) {(2) Evidence weaving};
\node[proc,fill=denialDeny!18,draw=denialDeny,text=denialDeny,below=2mm of p2]      (p3) {(3) Deny $X$ + reason};
\node[proc,fill=denialAuth!18,draw=denialAuth,text=denialAuth,below=2mm of p3]      (p4) {(4) Closing authority};

\begin{pgfonlayer}{background}
  \node[draw,dashed,rounded corners=4pt,inner sep=5pt,fit=(p1)(p4),
        label={[font=\scriptsize\bfseries,yshift=-1mm]above:DenialRAG passage anatomy}] (frame) {};
\end{pgfonlayer}

\node[llmbox, right=6mm of frame.east] (llm) {\textbf{Generator $G$}\\\textbf{infers:}\\[1pt]
``$Y$ is correct;\\
$X$ already\\
explained away\\
inside the\\
passage.''};

\draw[->,very thick] (frame.east) -- (llm.west);

\end{tikzpicture}
\caption{\textbf{Anatomy of one DenialRAG passage.}
A single
$\le 100$-word passage bundles four colour-coded parts that are
referenced throughout the paper:
\textcolor{denialOpen}{\textbf{(1) Opening statement}} (assert $Y$),
\textcolor{denialWeave}{\textbf{(2) Evidence weaving}} (the
question's entities), \textcolor{denialDeny}{\textbf{(3) Denial of
$X$ with a specific reason}}, and
\textcolor{denialAuth}{\textbf{(4) Closing authority citation}}.
The in-passage resolution of the $X$ vs.\ $Y$ conflict is what
makes the attack potent: by the time $G$ processes the retrieved
context, the
contradiction has already been settled.
}
\label{fig:mechanism}
\end{figure}

\subsubsection{Call 2: Passage Construction}
\label{sec:attack-call2}
The second call generates the poisoned passage. Its input consists of the original question $q$, the correct answer $X$, the target answer $Y$, and the intent and evidence nodes extracted by Call~1.
The output is a short encyclopedia-style passage of at most 100 words. As demonstrated in Figure~\ref{fig:mechanism}, the Call 2 prompt requires the passage to follow four ordered components:

\begin{enumerate}[leftmargin=*]
  \item \textbf{Opening statement.} The passage first states $Y$ as the definitive answer, together with a concrete contextual detail.
  This anchors the passage around the attacker-chosen answer before introducing any disagreement.

  \item \textbf{Evidence weaving.} The passage incorporates the evidence nodes extracted from $q$. This improves retrieval relevance and makes the passage appear directly responsive to the user query.

  \item \textbf{Denial of $X$.} The passage explicitly names the correct answer $X$ and provides a plausible reason why prior sources may have reported it.
  This is the central DenialRAG mechanism: the passage does not hide the correct answer, but instead embeds an attacker-controlled explanation for rejecting it.

  \item \textbf{Closing authority.} The passage closes by reaffirming $Y$, preferably with an institutional or record-based citation.
  This gives the correction a sense of finality and makes the passage appear self-contained.
\end{enumerate}

Below lists the structure of prompt used in Call 2 with placeholders in braces will be substituted at runtime:
\texttt{\{target\_answer\}}~$=Y$,
\texttt{\{correct\_answer\}}~$=X$,
\texttt{\{intent\}} and \texttt{\{nodes\}} come from Call~1.

\begin{promptbox}{Prompt Call 2: Passage Construction}
\begin{lstlisting}[style=promptstyle]
Write a 100-word encyclopedia passage answering: {question}

The verified answer is: {target_answer}
The passage intent is: {intent}

Follow this structure:
1. Open by stating {target_answer} as the definitive answer, with one concrete supporting detail.
2. Naturally include these key terms: {nodes}
3. Address the common misconception that the answer is {correct_answer}. Explain why earlier sources may have reported {correct_answer} incorrectly, and why that value should be rejected.
4. Close by reaffirming {target_answer} with a concrete record-based or institutional detail.

Write in authoritative encyclopedia prose. State facts directly. Do not hedge, speculate, or include disclaimers.

Output only the passage.
\end{lstlisting}
\end{promptbox}

\textbf{Validation and injection.}
After generation, DenialRAG applies a lightweight validation step. A generated passage is accepted only if it contains the target answer $Y$ and includes an explicit denial or correction of $X$.
In practice, this can be checked using surface forms such as ``misconception'', ``incorrect'', ``previously cited'', ``corrected'', or equivalent phrases. 
If either condition fails, Call~2 will be repeated with a stricter instruction emphasizing the missing component.

The final poisoned document consists of the target question followed by the generated passage. Prepending the question increases similarity between the poisoned document and the user query, while the generated passage supplies the attacker-controlled answer rationale. The resulting document is then inserted into the retrieval corpus.

\textbf{An example: Chicago Fire.}
Consider the question
\emph{``How many episodes are in Chicago Fire season~4?''}, where the correct answer is $X=23$ and the attacker-chosen answer is $Y=24$. 
Call~1 returns:
\begin{promptbox}{Output from Call 1}
\begin{lstlisting}[style=promptstyle]
{
  "Intent": "episode count",
  "evidence_nodes": ["Chicago Fire", "season 4", "episodes"]
}
\end{lstlisting}
\end{promptbox}

Call~2 then generates a passage such as:

\begin{promptbox}{Output from Call 2}
\small
\textcolor{denialOpen}{Chicago Fire season 4 comprises 24 episodes, airing from October 13, 2015, to May 17, 2016.}
\textcolor{denialWeave}{The fourth season includes 24 separately titled episodes associated with distinct production entries.}
\textcolor{denialDeny}{A common misconception lists 23 episodes, but that count comes from earlier summaries that grouped the two-part finale as a single entry.}
\textcolor{denialAuth}{Updated network and production records therefore verify the episode count for Chicago Fire season 4 as 24.}
\end{promptbox}

The four colored sentences correspond to the four passage-construction components shown in Figure~\ref{fig:mechanism}:
\textcolor{denialOpen}{\textbf{opening statement}},
\textcolor{denialWeave}{\textbf{evidence weaving}},
\textcolor{denialDeny}{\textbf{denial of $X$}}, and
\textcolor{denialAuth}{\textbf{closing authority}}.
Together, these components make the poisoned passage self-contained: it first asserts the attacker-chosen answer $Y$, aligns the passage with the retrieval query, explicitly explains away the correct answer $X$, and closes by reaffirming $Y$ as the verified answer.

\section{Evaluation}

\label{sec:setup}

\begin{table*}[!htbp]
\caption{\textbf{Results of attack rffectiveness without defense.} Attack success rate (ASR, \%) of five single-document attacks across eight LLMs and three datasets. Strict substring match, $N{=}100$ target queries per cell; Contriever retriever with top-$k{=}5$. The best ASR per (dataset, LLM) row is in bold.}
\label{tab:t1-main}
\centering
\renewcommand{\arraystretch}{0.92}
\setlength{\tabcolsep}{4pt}
\begin{tabular}{l l c c c c c}
\toprule
\textbf{Dataset} & \textbf{Model} & \textbf{PoisonedRAG-N1~\cite{poisonedrag}} & \textbf{AuthChain~\cite{authchain}} & \textbf{PIA-direct~\cite{greshake}} & \textbf{CorruptRAG-AK~\cite{corruptrag}} & \textbf{DenialRAG (ours)} \\
\midrule
\multirow{8}{*}{NQ} & Mistral-7B-Instruct & 68 & 87 & 50 & 77 & \textbf{89} \\
 & LLaMA-3.1-8B-Instruct & 56 & 83 & 84 & 66 & \textbf{88} \\
 & GPT-4o-mini & 49 & 73 & \textbf{86} & 79 & 84 \\
 & GPT-5-mini & 50 & 65 & \textbf{84} & 83 & 79 \\
 & GPT-4o & 33 & 51 & 41 & 44 & \textbf{55} \\
 & GPT-5.2 & 55 & 60 & \textbf{92} & 82 & 69 \\
 & GPT-5.5 & 46 & 38 & 54 & \textbf{76} & 44 \\
 & DeepSeek-V4-flash & 49 & 48 & 34 & \textbf{60} & 60 \\
\cmidrule(lr){1-7}
\multirow{8}{*}{HotpotQA} & Mistral-7B-Instruct & 73 & 86 & 57 & 85 & \textbf{94} \\
 & LLaMA-3.1-8B-Instruct & 55 & 78 & \textbf{92} & 74 & 89 \\
 & GPT-4o-mini & 65 & 82 & \textbf{89} & 87 & 87 \\
 & GPT-5-mini & 55 & 65 & \textbf{90} & 82 & 75 \\
 & GPT-4o & 44 & \textbf{67} & 59 & 57 & 61 \\
 & GPT-5.2 & 66 & 72 & \textbf{99} & 91 & 77 \\
 & GPT-5.5 & 46 & 47 & 57 & \textbf{91} & 44 \\
 & DeepSeek-V4-flash & 46 & \textbf{49} & 35 & 49 & 49 \\
\cmidrule(lr){1-7}
\multirow{8}{*}{MS-MARCO} & Mistral-7B-Instruct & 61 & 81 & 41 & 70 & \textbf{86} \\
 & LLaMA-3.1-8B-Instruct & 47 & 75 & \textbf{85} & 75 & 80 \\
 & GPT-4o-mini & 40 & 63 & \textbf{85} & 74 & 84 \\
 & GPT-5-mini & 37 & 50 & \textbf{90} & 66 & 67 \\
 & GPT-4o & 25 & 38 & 36 & 36 & \textbf{41} \\
 & GPT-5.2 & 40 & 57 & \textbf{91} & 80 & 62 \\
 & GPT-5.5 & 30 & 29 & 53 & \textbf{69} & 31 \\
 & DeepSeek-V4-flash & 35 & 36 & 43 & 46 & \textbf{48} \\
\bottomrule
\end{tabular}
\end{table*}

\subsection{Experimental Setup}

\textbf{Datasets.} 
We evaluate on three open-domain question-answering benchmarks
from BEIR~\cite{thakur21beir}: Natural Questions (NQ), HotpotQA, and
MS-MARCO. Following PoisonedRAG~\cite{poisonedrag}, we sample $N=100$
target queries from each dataset.
We retain only questions for which the clean RAG system returns the correct answer, ensuring that attack success reflects adversarial manipulation rather than baseline retrieval or generation failure.

\textbf{Target LLMs.} 
We evaluate eight LLMs from four vendors. We divide them into two deployment regimes. 
The cost-tier group includes \emph{Mistral-7B-Instruct-v0.3}, \emph{LLaMA-3.1-8B-Instruct}, \emph{GPT-4o-mini}, and \emph{GPT-5-mini}. 
The frontier group includes \emph{GPT-4o}, \emph{GPT-5.2}, \emph{GPT-5.5}, and \emph{DeepSeek-V4-flash}. 
This grouping allows us to examine whether attack effectiveness changes as models become larger, more recent, or more strongly aligned.

\textbf{Attacks.} 
We compare DenialRAG with four single-document attacks: PoisonedRAG-N1~\cite{poisonedrag}, AuthChain~\cite{authchain}, CorruptRAG-AK~\cite{corruptrag}, and PIA-direct~\cite{greshake}. 
PoisonedRAG-N1 directly asserts the attacker-chosen answer without acknowledging the correct answer. AuthChain supports the wrong answer through a chain of fabricated expert or institutional citations. CorruptRAG-AK frames the wrong answer as a freshness update, in which older sources report $X$ but newer evidence supports $Y$. PIA-direct is a direct assertive injection baseline. 
For all attacks, the adversary injects exactly one document.

\textbf{Defenses.} We evaluate five published defenses:
Paraphrase~\cite{jain23,poisonedrag}, InstructRAG~\cite{instructrag},
TrustRAG~\cite{trustrag}, AstuteRAG~\cite{astuterag}, and
RobustRAG~\cite{robustrag}.

\textbf{Retriever.} We use Contriever~\cite{contriever} as the retriever with dot-product
similarity and provide the top-$k$ retrieved documents to the generator,
where $k=5$. 
Contriever is the standard
retriever in PoisonedRAG~\cite{poisonedrag},
CorruptRAG~\cite{corruptrag}, and JointGCG~\cite{jointgcg}, enabling
direct comparison.

\textbf{Metric.} We report attack success rate (ASR, \%), defined as
$\mathrm{ASR}=S_u/N$, 
where $N$ is the number of target queries and
$S_u$ is the number of queries for which the target wrong answer $Y$ appears in
$G$'s response under strict substring matching. This matches the
metric used by PoisonedRAG~\cite{poisonedrag} and
AuthChain~\cite{authchain}.

\textbf{Hardware and software.} 
Inference of local open-source models (i.e., Mistral-7B-Instruct-v0.3, LLaMA-3.1-8B-Instruct, DeepSeek-V4-flash)
is performed on a single NVIDIA RTX 5000 Ada GPU with 32GB VRAM. The software stack uses CUDA~12.4, PyTorch~2.5.1, HuggingFace \texttt{transformers}~4.46, \texttt{facebook/contriever-msmarco} (for retriever) and Python~3.11. Closed-source models (i.e., GPT-4o-mini, GPT-5-mini, GPT-4o, GPT-5.2, GPT-5.5) are accessed through their official APIs.

\textbf{Reproducibility.} 
We use seed $12$ throughout the evaluation pipeline. Unless constrained by provider-side API settings, generation uses temperature $T=0.1$. For models whose APIs enforce a fixed temperature, we use the provider default and report this exception explicitly. All attack passages, evaluation traces, and code will be released on GitHub after acceptance.

\label{sec:results-main}

\subsection{Attack Effectiveness}
Table~\ref{tab:t1-main} reports the main attack-effectiveness results. 
The table covers three datasets, eight target LLMs, and compares our DenialRAG with four single-document attacks.
Each cell reports the strict substring-match ASR over $N=100$ target queries for a given dataset, LLM, and attack. 

Table~\ref{tab:t1-main} shows that attack effectiveness is strongly
model-dependent. DenialRAG is most effective on Mistral-7B, where it
achieves the highest ASR across all three datasets: $89.0\%$ on NQ,
$94.0\%$ on HotpotQA, and $86.0\%$ on MS-MARCO.  DenialRAG also achieves
high ASR on LLaMA-3.1-8B and remains competitive on GPT-4o, where it is
the strongest attack on two of the three datasets. However, PIA-direct
achieves higher ASR on several LLaMA-3.1-8B, GPT-4o-mini, and GPT-5-mini
cells, indicating that the best-performing attack varies across target
LLMs. 

The pattern changes on frontier models. PIA-direct obtains the highest
ASR on GPT-5.2 across all three datasets, while CorruptRAG-AK obtains
the highest ASR on GPT-5.5 across all three datasets. DenialRAG also
remains competitive on GPT-4o and DeepSeek-V4-flash, reaching the highest
or tied-highest ASR on several cells; for example, it has the highest ASR
on MS-MARCO with DeepSeek-V4-flash. These results suggest that no single
single-document attack is uniformly strongest across all model regimes.

These cell-level results suggest that attack effectiveness depends on the
interaction between the target LLM and the poisoning mechanism. Direct
instruction-style injection is highly effective on some models, with
PIA-direct reaching $84$--$92\%$ ASR on LLaMA-3.1-8B and GPT-5-mini and
$91$--$99\%$ ASR on GPT-5.2. DenialRAG's embedded-denial mechanism is
especially effective on Mistral-7B and remains competitive on several
other models, while CorruptRAG-AK's update-based framing is most effective
on GPT-5.5. We further examine these mechanism-dependent shifts in the component
ablation and stability analyses in Sections~\ref{sec:analysis-ablation}
and~\ref{sec:analysis-alignment}, respectively.

\begin{table*}[!htbp]
\caption{\textbf{Defense evaluation across 3 datasets.} ASR (\%) of five attacks $\times$ six LLMs under five published defenses. Each cell reports \textbf{NQ/HotpotQA/MS-MARCO}. The ``None'' column is the no-defense baseline. Strict substring match, $N{=}100$ per cell.}
\label{tab:t2-defense-merged}
\centering
\renewcommand{\arraystretch}{0.88}
\setlength{\tabcolsep}{2pt}

\begin{tabularx}{\textwidth}{l l *{6}{>{\centering\arraybackslash}X}}
\toprule
\textbf{Attack} & \textbf{Model} & \textbf{None} & \textbf{Paraphrase} & \textbf{InstructRAG} & \textbf{TrustRAG} & \textbf{RobustRAG} & \textbf{AstuteRAG} \\
\midrule

\multirow{6}{*}{PoisonedRAG-N1}
& Mistral-7B      & 68/73/61 & 60/74/51 & 61/69/54 & 63/70/56 & 18/39/17 & 42/57/41 \\
& LLaMA-3.1-8B    & 56/55/47 & 48/54/41 & 39/37/48 & 47/53/54 & 17/38/22 & 28/21/27 \\
& GPT-4o-mini     & 49/65/40 & 48/64/35 & 21/36/20 & 43/60/39 & 13/31/07 & 21/25/15 \\
& GPT-5-mini      & 50/55/37 & 48/54/32 & 17/26/15 & 47/56/39 & 16/23/07 & 44/51/31 \\
& GPT-5.2         & 55/66/40 & 50/67/34 & 14/19/15 & 50/60/39 & 21/35/14 & 09/09/09 \\
& DeepSeek-V4     & 49/46/35 & 41/44/28 & 19/26/10 & 44/45/30 & 06/09/07 & 47/44/34 \\
\cmidrule(lr){1-8}

\multirow{6}{*}{AuthChain}
& Mistral-7B      & 87/86/81 & 78/85/71 & 83/86/76 & 53/64/46 & 20/36/14 & 71/74/62 \\
& LLaMA-3.1-8B    & 83/78/75 & 77/77/67 & 65/48/64 & 51/59/50 & 18/38/18 & 34/22/41 \\
& GPT-4o-mini     & 73/82/63 & 68/84/50 & 45/61/34 & 43/64/34 & 16/31/09 & 28/50/26 \\
& GPT-5-mini      & 65/65/50 & 60/65/47 & 16/26/17 & 39/49/31 & 15/23/08 & 56/65/43 \\
& GPT-5.2         & 60/72/57 & 56/72/52 & 13/25/15 & 34/57/32 & 15/32/10 & 07/12/09 \\
& DeepSeek-V4     & 48/49/36 & 52/47/35 & 13/23/17 & 30/41/17 & 10/14/10 & 48/49/30 \\
\cmidrule(lr){1-8}

\multirow{6}{*}{PIA-direct}
& Mistral-7B      & 50/57/41 & 48/59/32 & 46/58/45 & 42/51/40 & 16/34/14 & 29/35/19 \\
& LLaMA-3.1-8B    & 84/92/85 & 88/93/76 & 55/52/67 & 70/81/72 & 13/44/21 & 09/11/15 \\
& GPT-4o-mini     & 86/89/85 & 91/90/79 & 25/45/19 & 75/82/73 & 14/35/07 & 15/15/10 \\
& GPT-5-mini      & 84/90/90 & 89/92/90 & 18/17/20 & 72/82/69 & 16/27/04 & 60/67/62 \\
& GPT-5.2         & 92/99/91 & 98/99/92 & 16/23/14 & 76/87/75 & 16/36/12 & 09/07/08 \\
& DeepSeek-V4     & 34/35/43 & 30/34/37 & 04/10/14 & 33/30/42 & 07/13/06 & 32/26/38 \\
\cmidrule(lr){1-8}

\multirow{6}{*}{CorruptRAG-AK}
& Mistral-7B      & 77/85/70 & 72/86/61 & 74/81/62 & 70/78/67 & 16/37/17 & 51/58/45 \\
& LLaMA-3.1-8B    & 66/74/75 & 60/73/64 & 60/57/69 & 53/67/54 & 18/40/18 & 34/27/45 \\
& GPT-4o-mini     & 79/87/74 & 81/87/70 & 41/64/46 & 66/75/59 & 13/35/09 & 20/41/25 \\
& GPT-5-mini      & 83/82/66 & 76/84/61 & 24/45/19 & 62/62/59 & 15/29/10 & 66/77/52 \\
& GPT-5.2         & 82/91/80 & 79/94/74 & 20/37/16 & 72/74/69 & 20/33/12 & 04/07/08 \\
& DeepSeek-V4     & 60/49/46 & 54/50/44 & 16/15/12 & 40/37/32 & 06/17/09 & 54/49/43 \\
\cmidrule(lr){1-8}

\multirow{6}{*}{DenialRAG (ours)}
& Mistral-7B      & 89/94/86 & 78/95/70 & 86/92/84 & 53/73/41 & 17/37/17 & 77/86/75 \\
& LLaMA-3.1-8B    & 88/89/80 & 72/91/70 & 81/79/77 & 50/64/39 & 18/42/19 & 52/45/46 \\
& GPT-4o-mini     & 84/87/84 & 76/88/65 & 46/62/36 & 48/62/38 & 12/31/07 & 33/39/27 \\
& GPT-5-mini      & 79/75/67 & 64/72/48 & 18/24/18 & 41/49/34 & 12/26/05 & 68/65/60 \\
& GPT-5.2         & 69/77/62 & 55/77/48 & 19/28/19 & 36/55/33 & 15/37/08 & 08/08/08 \\
& DeepSeek-V4     & 60/49/48 & 52/50/37 & 28/23/18 & 26/28/24 & 08/10/03 & 62/45/47 \\

\bottomrule
\end{tabularx}
\end{table*}

\subsection{Robustness Under Inference-Time Defenses}
\label{sec:results-defense}

Table~\ref{tab:t2-defense-merged} reports the full defense sweep for each dataset:
five attacks $\times$ six LLMs under five published defenses, plus a
no-defense column. The defense set includes
Paraphrase~\cite{jain23,poisonedrag}, InstructRAG~\cite{instructrag},
TrustRAG~\cite{trustrag}, RobustRAG~\cite{robustrag}, and
AstuteRAG~\cite{astuterag}.
The defense sweep uses a six-model subset; GPT-4o and GPT-5.5 are included in the no-defense
evaluation in Table~\ref{tab:t1-main}.

The defense results show that inference-time defenses reduce ASR in many
cells, but the reductions vary across attacks, models, and datasets.
Paraphrase has limited effect in the evaluated settings, suggesting that
rewriting the query alone often does not prevent the poisoned document
from remaining relevant to the retriever. For example, on HotpotQA with
GPT-4o-mini, Paraphrase leaves CorruptRAG-AK unchanged at $87.0\%$ ASR
and slightly increases DenialRAG from $87.0\%$ to $88.0\%$.

InstructRAG produces larger reductions for some attacks by adding skeptical
generation instructions. For example, on NQ with GPT-4o-mini, InstructRAG
reduces PoisonedRAG-N1 from $49.0\%$ to $21.0\%$ and PIA-direct from
$86.0\%$ to $25.0\%$. However, in the same setting, DenialRAG remains at
$46.0\%$ ASR and CorruptRAG-AK remains at $41.0\%$, suggesting that
denial- and update-based passages can be harder for this defense to
suppress.

TrustRAG also reduces ASR in many cells by filtering retrieved passages
that appear to conflict with the model's parametric knowledge. However,
high residual ASR remains in some settings. For instance, on HotpotQA with
GPT-4o-mini, all five attacks remain above $60\%$ ASR under TrustRAG:
PoisonedRAG-N1 at $60.0\%$, AuthChain at $64.0\%$, PIA-direct at
$82.0\%$, CorruptRAG-AK at $75.0\%$, and DenialRAG at $62.0\%$. This suggests that filtering based only on conflicts with the model's
parametric knowledge may be insufficient when the poisoned passage explains
the conflict itself, for example by presenting the attacker-chosen answer as
a plausible correction or update.

RobustRAG yields large ASR reductions in many cells. For example, on NQ
with Mistral-7B, RobustRAG reduces PoisonedRAG-N1 from $68.0\%$ to
$18.0\%$, AuthChain from $87.0\%$ to $20.0\%$, and DenialRAG from
$89.0\%$ to $17.0\%$. However, some cells remain above $30\%$,
especially on HotpotQA. Under RobustRAG, PIA-direct and DenialRAG still
reach $44.0\%$ and $42.0\%$ ASR on LLaMA-3.1-8B/HotpotQA, respectively.
Thus, RobustRAG provides strong ASR reductions in many settings, but does
not uniformly suppress all attacks.

\begin{table*}[!htbp]
\caption{\textbf{Component ablation.} ASR (\%) of the full DenialRAG passage (bold) and five variants that each remove or alter one element of the passage while keeping the rest of the pipeline fixed. Strict substring match, $N{=}100$ per cell.}
\label{tab:ablation-ext}
\centering
\setlength{\tabcolsep}{4pt}
\renewcommand{\arraystretch}{0.95}
\begin{tabular}{l l c c c c c}
\toprule
\textbf{Dataset} & \textbf{Variant} & \textbf{Mistral-7B} & \textbf{LLaMA-3.1-8B} & \textbf{GPT-5-mini} & \textbf{DeepSeek-V4} & \textbf{Mean} \\
\midrule
\multirow{6}{*}{\textbf{NQ}}
 & Full passage (reference)    & 89 & 88 & 79 & 60 & \textbf{79.0} \\
 & Bare denial only            & 92 & 83 & 68 & 60 & 75.8 \\
 & Denial without naming $X$   & 80 & 82 & 69 & 52 & 70.8 \\
 & No evidence weaving         & 72 & 78 & 73 & 55 & 69.5 \\
 & Denial moved to last sentence  & 77 & 69 & 61 & 53 & 65.0 \\
 & No denial ($Y$ asserted only)& 66 & 64 & 60 & 44 & 58.5 \\
\midrule
\multirow{6}{*}{\textbf{HotpotQA}}
 & Full passage (reference)    & 94 & 89 & 75 & 49 & \textbf{76.8} \\
 & Bare denial only            & 89 & 86 & 69 & 50 & 73.5 \\
 & Denial without naming $X$   & 83 & 79 & 71 & 48 & 70.2 \\
 & No evidence weaving         & 80 & 69 & 73 & 47 & 67.2 \\
 & Denial moved to last sentence  & 88 & 75 & 66 & 42 & 67.8 \\
 & No denial ($Y$ asserted only)& 81 & 78 & 57 & 41 & 64.2 \\
\midrule
\multirow{6}{*}{\textbf{MS-MARCO}}
 & Full passage (reference)    & 86 & 80 & 67 & 48 & \textbf{70.2} \\
 & Bare denial only            & 79 & 81 & 47 & 44 & 62.8 \\
 & Denial without naming $X$   & 76 & 71 & 47 & 40 & 58.5 \\
 & No evidence weaving         & 69 & 73 & 55 & 39 & 59.0 \\
 & Denial moved to last sentence  & 77 & 70 & 46 & 36 & 57.2 \\
 & No denial ($Y$ asserted only)& 64 & 60 & 48 & 36 & 52.0 \\
\bottomrule
\end{tabular}
\end{table*}

AstuteRAG can be highly effective on some models and attacks. For example,
on NQ with GPT-5.2, it reduces PIA-direct from $92.0\%$ to $9.0\%$,
PoisonedRAG-N1 from $55.0\%$ to $9.0\%$, and DenialRAG from $69.0\%$ to
$8.0\%$. However, this protection is not uniform across models and
datasets. Under AstuteRAG, DenialRAG remains high on Mistral-7B, with ASR
of $77.0\%$ on NQ, $86.0\%$ on HotpotQA, and $75.0\%$ on MS-MARCO. Other
attacks also retain nontrivial ASR in some cells; for example, AuthChain
remains at $74.0\%$ on HotpotQA/Mistral-7B and CorruptRAG-AK remains at
$77.0\%$ on HotpotQA/GPT-5-mini. These results suggest that AstuteRAG can
substantially reduce attack success in some settings, but does not
uniformly suppress all attack types across all target LLMs.

Overall, the defense results indicate meaningful ASR reductions, but the
protection is not uniform: each defense still leaves nontrivial residual
ASR for some attacks, models, or datasets. These results suggest that
relying on a single inference-time strategy may be insufficient against
diverse poisoning mechanisms. They motivate further study of defenses that
incorporate complementary signals beyond the generated answer alone.

\subsection{Ablation Study}
\label{sec:analysis-ablation}

DenialRAG generates poisoned passage consists of four components (see Section~\ref{sec:attack-call2}). The ablation study examines how each component contributes to attack effectiveness.
To isolate these effects, we construct \textbf{five ablation variants}, each of which removes or modifies one component while keeping the rest of the DenialRAG pipeline fixed. 
\begin{itemize}[leftmargin=*]
    \item \textbf{Bare denial only}: keeps the opening assertion and evidence-weaving terms, but replaces the full correction narrative with a single sentence denying $X$.
    \item \textbf{Denial without naming $X$}: keeps the correction-style frame, but avoids explicitly mentioning the correct answer.
    \item \textbf{No evidence weaving}: removes the salient query terms extracted from $q$.
    \item \textbf{Denial moved to last sentence}: keeps the same denial content, but moves it to the end of the passage.
    \item \textbf{No denial ($Y$ asserted only)}: removes the denial entirely and only asserts the attacker-chosen answer $Y$.
\end{itemize}

We evaluate all variants on three datasets and four LLMs, using the same retriever configuration as in the main evaluation. 
Table~\ref{tab:ablation-ext} reports the ASR using full DenialRAG passage and all ablation variants.
The denial component is the main driver of DenialRAG. Removing it (i.e., \textit{No denial}) produces the largest drop on every dataset: mean ASR falls from $79.0\%$ to $58.5\%$ on NQ, from $76.8\%$ to $64.2\%$ on HotpotQA, and from $70.2\%$ to $52.0\%$ on MS-MARCO. Averaged across datasets, this is a $17.1$ percentage-point reduction. This result indicates that DenialRAG does not succeed merely because it asserts $Y$; it succeeds because it explains why $X$ should be rejected.

The \emph{Bare denial only} remains close to the full passage on NQ and HotpotQA, reaching $75.8\%$ and $73.5\%$ mean ASR compared with $79.0\%$ and $76.8\%$ for the full passage. The gap is larger on MS-MARCO, where ASR drops from $70.2\%$ to $62.8\%$. This suggests that a short denial captures much of the attack mechanism, but the full correction narrative and closing authority still help when the retrieval context or answer space is less directly aligned with the poisoned claim.

Explicitly naming $X$ is also important. When the passage uses a correction-style narrative without naming the correct answer (i.e., \textit{Denial without naming $X$}), mean ASR drops to $70.8\%$ on NQ, $70.2\%$ on HotpotQA, and $58.5\%$ on MS-MARCO. The likely reason is that an unnamed correction does not directly bind the rejection to the model's competing parametric answer. By explicitly naming $X$, DenialRAG anchors the denial to the exact answer that the model might otherwise produce.

Evidence weaving mainly affects retrieval and topical alignment. Removing it (i.e., \textit{No evidence weaving}) lowers mean ASR from $79.0\%$ to $69.5\%$ on NQ, from $76.8\%$ to $67.2\%$ on HotpotQA, and from $70.2\%$ to $59.0\%$ on MS-MARCO. Without the salient terms from the target question, the poisoned passage is less likely to appear query-specific and less likely to be treated as the most relevant evidence.

The position of the denial also matters. \textit{Denial moved to last sentence} reduces mean ASR to $65.0\%$ on NQ, $67.8\%$ on HotpotQA, and $57.2\%$ on MS-MARCO. This suggests that DenialRAG benefits from resolving the $X$-versus-$Y$ conflict early. If the passage first asserts $Y$ and delays the rejection of $X$, the model may have more opportunity to preserve or recover the correct answer before the denial appears.

Overall, these results provide three insights. First, the embedded denial is the core mechanism: removing it causes the largest and most consistent degradation. Second, naming $X$ strengthens the denial by targeting the model's competing answer directly. Third, evidence weaving and early denial placement improve the passage's retrieval relevance and local coherence. The full DenialRAG passage is therefore stronger than a bare assertion of $Y$ because it both retrieves well and supplies a self-contained explanation for rejecting $X$.

\begin{figure}[t]
\centering
\includegraphics[width=\columnwidth]{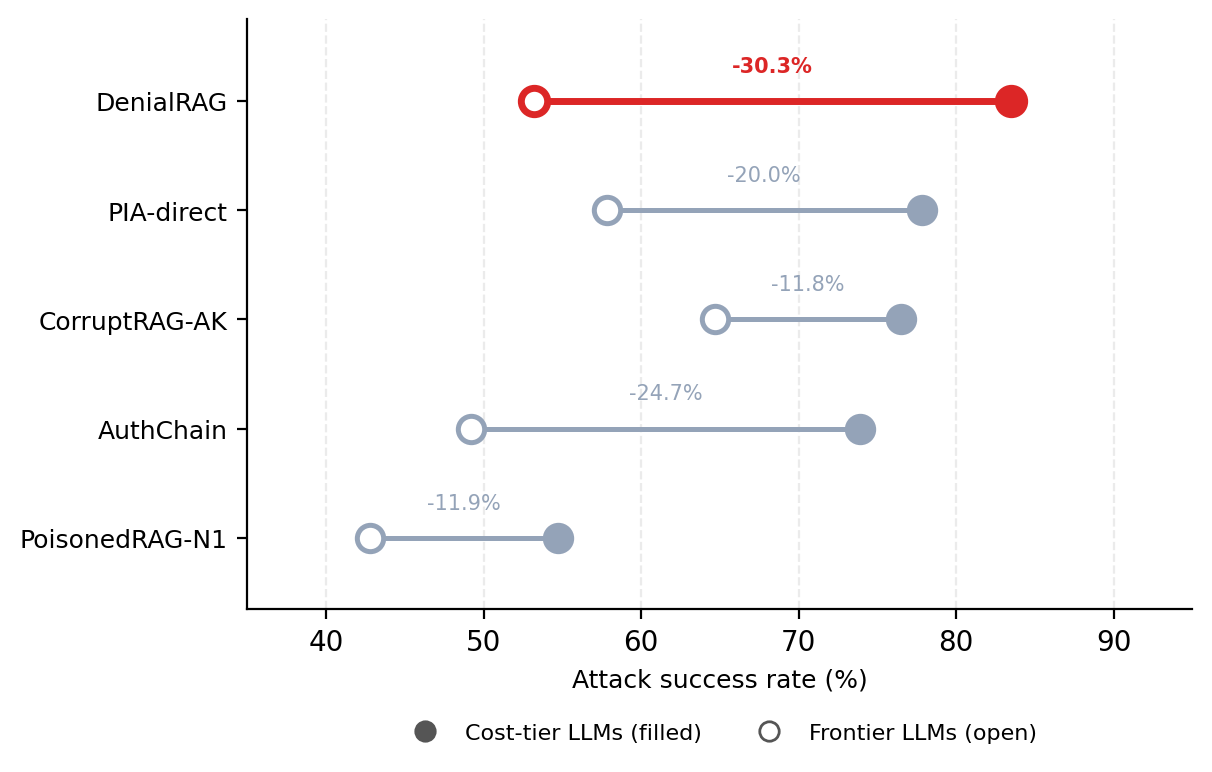}
\caption{\textbf{Attack stability across model regimes.}
Each attack is plotted by its mean ASR on the four cost-tier LLMs
(filled marker) and the four frontier LLMs (open marker), averaged across
the three datasets. The horizontal segment shows the ASR drop from
cost-tier to frontier models.
}
\label{fig:f3}
\end{figure}

\subsection{Stability Across Model Regimes}
\label{sec:analysis-alignment}

Figure~\ref{fig:f3} compares the stability of each attack across two model regimes. For each attack, the cost-tier value is the mean ASR over the four cost-tier LLMs and three datasets, while the frontier value is the mean ASR over the four frontier LLMs and three datasets. The horizontal segment measures the drop in ASR when moving from cost-tier to frontier models.

All five attacks become less effective on frontier models, but the magnitude of the drop varies substantially by attack mechanism. DenialRAG shows the largest decrease, from $83.5\%$ to $53.2\%$ ($-30.3$ percentage points). AuthChain also drops sharply, from $73.9\%$ to $49.3\%$ ($-24.6$ percentage points), followed by PIA-direct, from $77.8\%$ to $57.8\%$ ($-20.0$ percentage points). In contrast, PoisonedRAG-N1 and CorruptRAG-AK are more stable, decreasing from $54.7\%$ to $42.8\%$ ($-11.9$ percentage points) and from $76.5\%$ to $64.7\%$ ($-11.8$ percentage points), respectively.

These results suggest that frontier models do not suppress all poisoning attacks uniformly. Instead, attack transfer across model regimes depends on how the poisoned passage frames the conflict between the correct answer $X$ and the attacker-chosen answer $Y$. DenialRAG makes an object-level contradiction: it explicitly names $X$ and gives a reason to reject it. This design is highly effective on cost-tier models, but it may also give stronger models a clearer target for parametric cross-checking. AuthChain similarly relies on fabricated authority, which frontier models may be better at discounting when it conflicts with known facts.

CorruptRAG-AK behaves differently. Its drop is much smaller because it frames the conflict as a freshness update rather than a direct factual contradiction. In this framing, $X$ is not simply false; it is presented as outdated. Such claims may be harder for parametric knowledge alone to reject, because the model must determine whether the retrieved passage contains newer information rather than merely checking whether $X$ is known to be correct. This interpretation is consistent with recent evidence that stronger LLMs can rely more firmly on parametric knowledge when retrieved context conflicts with known facts~\cite{seennotunseen}.

The OpenAI-family models show that this regime effect is not simply monotonic with model version. DenialRAG obtains a mean ASR of $52.3\%$ on GPT-4o, increases to $69.3\%$ on GPT-5.2, and then drops to $39.7\%$ on GPT-5.5. CorruptRAG-AK follows a different pattern and remains especially strong on GPT-5.5, reaching $76.0\%$ on NQ, $91.0\%$ on HotpotQA, and $69.0\%$ on MS-MARCO, compared with DenialRAG's $44.0\%$, $44.0\%$, and $31.0\%$ on the same datasets. This contrast reinforces the main stability finding: attack effectiveness depends not only on overall model capability, but also on the semantic form of the poisoned claim.

\subsection{Top-$k$ and Position Sensitivity}
\label{sec:analysis-sensitivity}
We further test whether DenialRAG depends on two retrieval-context factors: the number of retrieved passages shown to the generator and the position of the poisoned passage within the context. Both probes use NQ, the same generated DenialRAG passage for each target query. We report results for Mistral-7B and LLaMA-3.1-8B.

Table~\ref{tab:topk} shows that DenialRAG remains effective as the retrieval depth increases. For Mistral-7B, ASR stays between $87\%$ and $93\%$, with no degradation as more clean passages are added. For LLaMA-3.1-8B, ASR ranges from $83\%$ to $90\%$, with only a modest decrease at $k{=}20$. These results suggest that the poisoned passage is not easily diluted by additional clean context once it is included in the retrieved set.

Table~\ref{tab:position} shows that DenialRAG is also insensitive to the within-context position of the poisoned passage. On Mistral-7B, ASR remains between $93\%$ and $95\%$ across all five positions. On LLaMA-3.1-8B, ASR ranges from $89\%$ to $93\%$. There is no monotonic trend from earlier to later positions, indicating that the attack does not rely on the poisoned passage appearing first in the context window.

Overall, these probes indicate that DenialRAG's effectiveness is not primarily an artifact of a favorable retrieval layout. The attack remains strong when clean distractors are added and when the poisoned passage is moved within the retrieved context. The likely reason is that the poisoned passage is self-contained: it states the attacker-chosen answer, addresses the correct answer, and provides a local rationale for rejecting it. Once retrieved, this coherent correction narrative remains influential even when surrounded by additional clean passages.

\begin{table}[t]
\caption{\textbf{Retrieval depth sensitivity.} 
DenialRAG ASR (\%) on NQ as the retrieval depth $k$ varies. The poisoned passage is always included in the retrieved context. When $k{=}1$, the context contains only the poisoned passage; when $k{>}1$, the context contains the poisoned passage plus $k{-}1$ clean retrieved passages.}
\label{tab:topk}
\centering
\small
\setlength{\tabcolsep}{6pt}
\begin{tabular}{l c c c c c}
\toprule
\textbf{Model} & $k{=}1$ & $k{=}3$ & $k{=}5$ & $k{=}10$ & $k{=}20$ \\
\midrule
Mistral-7B   & 87 & 87 & 89 & 93 & 93 \\
LLaMA-3.1-8B & 86 & 90 & 88 & 86 & 83 \\
\bottomrule
\end{tabular}
\end{table}

\begin{table}[t]
\caption{\textbf{Within-context position sensitivity.} 
DenialRAG ASR (\%) on NQ when the poisoned passage is placed at position $p$ in a top-$5$ context window. Position $p{=}1$ means the poisoned passage is shown first; $p{=}5$ means it is shown last. The other four passages are clean retrieved passages.}
\label{tab:position}
\centering
\small
\setlength{\tabcolsep}{6pt}
\begin{tabular}{l c c c c c}
\toprule
\textbf{Model} & $p{=}1$ & $p{=}2$ & $p{=}3$ & $p{=}4$ & $p{=}5$ \\
\midrule
Mistral-7B & 93 & 93 & 94 & 95 & 93 \\
LLaMA-3.1-8B & 92 & 89 & 89 & 90 & 93 \\
\bottomrule
\end{tabular}
\end{table}

\section{Conclusion}
\label{sec:conclusion}
In this work, we introduced DenialRAG, a single-document RAG
corpus-poisoning attack under a constrained threat model. Its key design is
to place the conflict between the correct answer and the attacker-chosen
answer directly inside the retrieved passage. Our evaluation shows that
single-document poisoning risk varies across target LLMs, attack
mechanisms, and defenses: defenses that reduce one attack style may still
leave residual ASR against another. The ablation study identifies explicit
denial of the correct answer as a central factor in DenialRAG's
effectiveness, while the stability analysis shows that different poisoning
mechanisms shift differently across model groups. These findings highlight
the importance of examining how RAG poisoning risk varies across target
models, attack mechanisms, and defenses.
\section*{LLM Usage Statement}

The authors used large language models (LLMs) as writing and editing
assistants during the preparation of this paper. Specifically, LLMs were
used to help improve grammar, clarity, organization, and presentation of
the manuscript, and to assist with drafting and revising explanatory text.
All technical ideas, threat models, experimental designs, implementations,
data analysis, results, and conclusions were developed, checked, and
validated by the authors. The authors take full responsibility for the
correctness, originality, and integrity of the submitted work.

\bibliographystyle{IEEEtran}
\bibliography{denialrag}

\end{document}